\documentclass[a4paper,usenatbib]{mn2e}
\usepackage{natbib,epsfig}
\usepackage{xspace}

\catcode`\@=11
\def\gta{\ifmmode{\,\mathrel{\mathpalette\@versim>\,}}
    \else{$\,\mathrel{\mathpalette\@versim>}\,$}\fi}
\def\lta{\ifmmode{\,\mathrel{\mathpalette\@versim<\,}}
    \else{$\,\mathrel{\mathpalette\@versim<}\,$}\fi}
\def\@versim#1#2{\lower 2.9truept \vbox{\baselineskip 0pt \lineskip
    0.5truept \ialign{$\m@th#1\hfil##\hfil$\crcr#2\crcr\sim\crcr}}}
\catcode`\@=12  
\def\figref#1{Fig.~\ref{#1}}

\def\kms{\,{\rm km}\,{\rm s}^{-1}}

\def\pc{\,{\rm pc}}
\def\kpc{\,{\rm kpc}}

\newcommand{\vect}[1]{{\bf #1}}
\newcommand{\vecthat}[1]{\hat\vect{#1}}
\newcommand{\eqref}[1]{Eq. \ref{#1}}
\def\vot{\vect{v}_{s}}
\newcommand{\smt}{\sigma_\mu}
\newcommand{\svot}{\sigma_{\vot}}
\newcommand{\cov}{{\rm cov}}

\newcommand{\votx}{v_{s, x}}
\newcommand{\voty}{v_{s, y}}
\newcommand{\vpar}{v_{s\parallel}}
\newcommand{\vper}{v_{s\perp}}
\def\svper{\sigma_{\vper}}
\def\svpar{\sigma_{\vpar}}

\def\masyr{\,{\rm mas}\,{\rm yr}^{-1}}
\def\deg{^{\circ}}

\newcommand{\krh}{K09\xspace}
\def\vo{\vect{v}_0}
\def\near{\sim\!}

\def\tx{t_\parallel}
\def\ty{t_\perp}
\def\rperi{R_{\rm p}}
\def\paper1{Paper~I\xspace}
\def\angwid{\Delta \psi}
\def\anglen{\Delta \theta}
\def\lendeproj{\Delta \Theta}

\title[Galactic parallax of the tidal stream GD-1]
{An application of Galactic parallax: the distance to the
tidal stream GD-1}

\author[A. Eyre]{Andy Eyre\\
Rudolf Peierls Centre for Theoretical Physics, Keble Road, Oxford OX1 3NP, UK\\}

\begin{document}

\date{17 October 2009}

\pagerange{\pageref{firstpage}--\pageref{lastpage}} \pubyear{2009}

\maketitle

\label{firstpage}

\begin{abstract}
We assess the practicality of computing the distance to stellar streams
in our Galaxy, using the method of Galactic parallax suggested
by \cite{paper3}. We find that the uncertainty in Galactic parallax
is dependent upon the specific geometry of the problem in question.
In the case of the tidal stream GD-1, the problem geometry indicates
that available proper motion data, with individual accuracy $\sim4\masyr$,
should allow
estimation of its distance with about 50 percent uncertainty.
Proper motions accurate to $\sim1\masyr$, which are expected from the forthcoming
Pan-STARRS PS-1 survey, will allow estimation of its distance to about 10 percent
uncertainty. Proper motions from the future LSST and Gaia projects will be more accurate still,
and will allow the parallax for a stream $30\kpc$ distant to be measured with
$\near 14$ percent uncertainty.

We demonstrate the feasibility of the method and show that our
uncertainty estimates are accurate by computing Galactic parallax
using simulated data for the GD-1 stream. We also apply the method
to actual data for the GD-1 stream, published by \cite{koposov}.
With the exception of one datum, the distances estimated using
Galactic parallax match photometric estimates with less
than $1\kpc$ discrepancy. The scatter in the distances recovered using Galactic parallax
is very low, suggesting that the proper motion uncertainty reported
by \cite{koposov} is in fact over-estimated.

We conclude that the GD-1 stream is $(8\pm 1)\kpc$ distant, on
a retrograde orbit inclined $37\deg$ to the plane,
and that the visible portion of the stream is likely to be near pericentre.
\end{abstract}


\section{Introduction}

Measuring distances in our Galaxy is critical to
understanding its structure. However, line-of-sight
distances can typically be measured with only relatively
poor precision. This lack of precision is manifest
in the most basic of Galactic parameters; the solar
radius $R_0$ is hardly known to better than 5 percent
uncertainty \citep{gillessen}, and this result renders the circular
velocity at $R_0$ similarly uncertain \citep{mb09}. 
An accurate knowledge of distances is essential
to create convincing models of the Milky Way,
which in turn influence our understanding of
the physics of galaxy formation in general.

Conventional trigonometric parallax has long been used to calculate
accurate distances to nearby stars. The regular nature of the
parallactic motion of a star, caused by the Earth's orbit around the Sun, allows
this motion to be decoupled from the intrinsic proper motion of the
star in the heliocentric rest frame. Hence the distance to the
star can be calculated. However, the maximum
 baseline generating such parallaxes is obviously limited to 2AU. For
a given level of astrometric precision, this imposes a fundamental
limit to the observable distance. Indeed, the accuracy of
parallaxes reported by the Hipparcos mission
data \citep{newhipparcos} falls to $20$--$30$ percent at best for distances
$\near 300\pc$ and only then for the brightest stars.
Upcoming astrometric projects such
as Pan-STARRS \citep{panstarrs}, LSST \citep{lsst} and the
Gaia mission \citep{gaia} will achieve similar uncertainty
for Sun-like stars as distant as a few kpc, and at
fainter magnitudes than was possible with Hipparcos. This
extended range will encompass less than 1 percent of the total number
of such stars in our Galaxy.

It is clear that it will {\em not} soon be possible to calculate
distances to many of the stars in our Galaxy with conventional
trigonometric parallaxes. Alternative means to compute distances to
stars are therefore required. Photometry can be used to estimate the
absolute magnitude of a star which, when combined with its observed
magnitude, allows its distance to be computed. Unfortunately, all
attempts to calculate such photometric distances are hindered by the
same problems: obscuration by intervening matter alters both observed
magnitude \citep{vergely} and colour \citep{reddening1,drimmel}, and
it is difficult to model appropriate corrections without a reference
distance scale. The effects of chemical composition and age further
complicate matters \citep{juric}. It is therefore difficult to compute
photometric distances with an accuracy much better than 20 percent,
even for nearby stars, and distances to faint stars are less accurate
still \citep{juric}.

Latterly, it was realised \citep[hereafter \paper1]{paper3} that the orbital
motion of the Sun about the Galaxy could be used to
compute trigonometric distances to stars.
In the general case, it is not possible to
do this because the parallactic motion of 
a star and its intrinsic proper motion are
inextricably mixed up.
However, in the special case where the star can be associated
with a stellar stream, its rest-frame trajectory can be predicted
from the locations of the other associated stars.
Using this trajectory, the proper motion in the Galactic
rest frame can indeed be decoupled from the reflex
motion of the Sun, and the component of 
its motion due to parallax can be computed.

Such `Galactic parallaxes' have the same geometrical basis as
conventional trigonometric parallaxes, and as such as free from errors
induced by obscuration and reddening. However, the range of Galactic
parallax significantly exceeds that of conventional parallax.  This is
because the Sun orbits about the Galactic centre much faster than the
Earth orbits the Sun, and because, unlike with conventional parallax,
the Galactic parallax effect is cumulative with
continued observation. In realistic cases, for a typically oriented
stream, we can expect the Galactic parallax to be observable at nearly
40 times the distance of the equivalent trigonometric parallax, based
on 3 years of observations. For increased range, one simply observes over
a longer baseline.

This large range means that Galactic parallax might prove
a powerful tool to complement conventional parallaxes, and
validate other distance measuring tools. It is exciting to note that the
capabilities of astrometric projects such as LSST, which
will observe the conventional parallax of a G star at a distance of $\near 1\kpc$
with 20 percent uncertainty, will put much of the Galaxy in range of
Galactic parallax calculations with similar accuracy.

The main restriction on the use of Galactic parallax is the
requirement for stars to be part of a stream.  However, the continuing
discovery of significant numbers of streams
\citep{Odenkirchen02,sagdf,yanny,Fieldstars,gostream,GD-1,ngc5466,
g09,newberg} using optical
surveys implies that they are a staple feature of the Galactic
environment, rather than a rarity.  The deep surveys of Pan-STARRS and
LSST are likely to find yet more, increasing the 
number of applications for Galactic parallax.

This paper explores the viability of using Galactic parallax to
estimate distances and demonstrates its practicality by applying it to
data for the GD-1 stream \citep{GD-1} published by \citet[hereafter
\krh]{koposov}; we choose to work with the latter over the
earlier analysis of the same stream by \cite{willett} on account
of the significantly smaller proper motion uncertainties ($1\masyr$
vs $4\masyr$) cited in the later work.
Throughout this paper, the Solar motion is assumed to be $(U,V,W)
= (10.0,252,7.1) \pm (0.3,11,0.34)\kms$, consistent with \cite{ab09},
\cite{reid2004} and \cite{gillessen}.

The paper is arranged as follows.
Section \ref{sec:galacticparallax}
reviews the calculation of Galactic parallax,
and explores the uncertainty affecting such calculations,
and how this uncertainty affects practical application.
Section \ref{sec:tests} demonstrates the viability of the method
by applying it to pseudo-data, 
and Section \ref{sec:gd1} applies the method to actual data for the
GD-1 stream. Section \ref{sec:conclusions}
summarizes our conclusions.

\section{Galactic parallax} \label{sec:galacticparallax}

Suppose that a star is part of a stellar stream, and has a
location relative to the Sun described by $(\vect{x}-\vect{x_0})
= r\vecthat{r}$, where $r$ is the distance to the star, and
$\vect{x_0}$ is the position of the Sun.
In the plane of the sky, let the tangent to
the trajectory of the stream, near the star,
be indicated by the vector $\vecthat{p}$. Assume the
velocity of the Sun, $\vect{v}_0$, in the Galactic rest frame (grf) 
is known. \paper1 showed that if the measured proper motion of
the star is $\mu \vecthat{t}$, then
\begin{equation}
\dot{u}\vecthat{p} = \mu \vecthat{t} + \frac{\vot}{r}
= \mu \vecthat{t} + \Pi \vot,
\label{eq:fundamental}
\end{equation}
where $\Pi \equiv 1/r$ is the Galactic parallax,
$\dot{u}$ is the proper motion as would be seen from the grf,
and $\vot$ is the Sun's velocity projected into the plane
of the sky. We note that $\dot{u} = v_t/r$, where $v_t$ is that
component of the star's grf velocity perpendicular to the line-of-sight,
and that,
\begin{equation}
\vot = (\vect{v}_0 - \vecthat{r} \cdot \vect{v}_0\, \vecthat{r} ).
\end{equation}
\eqref{eq:fundamental} is a vector expression and can be solved
simultaneously for both $\dot{u}$ and $\Pi$ provided that $\vecthat{p}$,
$\vecthat{t}$ and $\vot$ are not parallel.
The stream direction $\vecthat{p}$ will not typically be
known outright, but must be estimated from the positions of
stream stars on the sky. We can achieve this by fitting a low order
curve through the position data, the tangent of which is then taken
to be $\vecthat{p}$. The curve must be
chosen to reproduce the gross behaviour of the stream,
but we must avoid fitting high-frequency noise, because $\vecthat{p}$ is
a function of the derivative of this curve, which is sensitive
to such noise.

\subsection{Uncertainty in Galactic parallax calculations}

We begin by rendering \eqref{eq:fundamental} into
an orthogonal on-sky coordinate system, whose
components are denoted by $(x,y)$.
In this coordinate system \eqref{eq:fundamental}
can be solved for $\Pi$,
\begin{equation}
\Pi = { {\mu \left(t_x \sin \alpha - t_y \cos \alpha\right)} \over
{\voty \cos \alpha- \votx \sin \alpha} }, \label{eq:pi1}
\end{equation}
where the $(x,y)$ suffixes denote the corresponding components of
their respective vectors, and where we have defined the
angle $\alpha \equiv \arctan (p_y/p_x)$.

The choice of coordinates $(x,y)$ is arbitrary.
We are therefore free to choose the coordinate system in which
$\alpha = 0$, i.e. that system in which the $x$-axis points
along the stream trajectory, $\vecthat{p}$. \eqref{eq:pi1} becomes,
\begin{equation}
\Pi = -{\mu \ty \over \vper}, \label{eq:pi2}
\end{equation}
where we now identify the $y$-component of the various vectors
as that component perpendicular ($\perp$) to the stream trajectory,
and the $x$-component as that component parallel ($\parallel$) to
the trajectory. \eqref{eq:pi2} shows explicitly that the Galactic
parallax effect is due to the reflex motion of stream stars perpendicular
to the direction of their travel.

Uncertainties in the $(x,y)$ components of the measured quantities
$\mu \vecthat{t}$ and $\vot$, and uncertainty in $\alpha$,
can be propagated to $\Pi$ using \eqref{eq:pi1}. When we set
$\alpha = 0$, this equation becomes,
\begin{equation}
{\sigma^2_\Pi \over \Pi^2} =
{\smt^2 \over \mu^2 \ty^2} + 
{\svper^2 \over \vper^2}
+ {\sigma_\alpha^2 \over \vper^2}\left(\vpar + {\mu \tx \over \Pi} \right)^2,
\label{eq:error:pi}
\end{equation}
where we anticipate the uncertainty in $\mu\vecthat{t}$ to be
isotropic, and so we have set $\sigma_{\mu t_x}= \sigma_{\mu t_y} =
\smt$.

We assume $\smt$ to be known from observations; it may contain
any combination of random and systematic error.
$\sigma_{\vper}$ is calculated directly from
the error ellipsoid on $\vect{v}_0$, which is assumed known. Any
error on $\vo$ affects all data in exactly the same way. However, the
projection of error on $\vo$ to $\vot$ varies with position on the
sky. Hence, the effect of $\svot$ is to produce a systematic error in
reported distance that varies along the stream in a problem-specific
way.

Uncertainty in $\alpha$ arises from two sources.  Firstly, because
the on-sky trajectory $\vecthat{p}$ is chosen by fitting a smooth
curve through observational fields, $\vecthat{p}$ need not be
exactly parallel to the underlying stream.
Further, since $\vecthat{p}$ depends on the derivative of the
fitted curve, it is likely to be much less well
constrained for the data points at the ends of the stream than for
those near the middle.

We can quantify this effect. At the endpoints, the fitted curve is likely
to depart from the stream by at most $\angwid$, the angular width of the stream
on the sky. For a low-order curve, this departure is likely to have been gradual over
approximately half the angular stream length, $\anglen$, giving a contribution to
$\sigma_\alpha$ from fitting of,
\begin{equation}
\sigma^2_{\alpha,f} = {4\angwid^2 \over \anglen^2}.
\label{eq:errora:curvefitting}
\end{equation}

The second contribution to $\sigma_\alpha$ arises as follows.
Since the stream has finite width, at any point, the stars within it
have a spread of velocities, corresponding to the spread
in action of the orbits that make up the stream.
If the stars in a stream show a spread in velocity $(\sigma_{v_x}, \sigma_{v_y})$
about a mean velocity $\vect{v_t} = r\dot{u}\vecthat{p}$, this effect
contributes,
\begin{equation}
\sigma^2_{\alpha,v} = {1\over v_t^2}\left( \sigma_{v_y}^2 \cos^2 \alpha +
\sigma_{v_x}^2 \sin^2 \alpha \right),
\label{eq:alpha:vspread}
\end{equation}
to the uncertainty in $\alpha$ for a single star.
Again we can choose $\alpha = 0$, such that $\sigma_{v_y}
= \sigma_{v\perp}$, the velocity dispersion perpendicular to the stream direction.
\eqref{eq:alpha:vspread} becomes,
\begin{equation}
\sigma^2_{\alpha,v} = {\sigma^2_{v\perp} \over v_t^2} = {\sigma^2_{v\perp} \over 
(v \sin \beta)^2},
\label{eq:alpha:sigw}
\end{equation}
where we have introduced $v$, the grf speed of the stream, and
$\beta$, the angle of the stream to the line-of-sight.
$\sigma_{v\perp}$ has its origin in the random
motions of stars that existed within the progenitor object.
In fact, if we assume the stream has not spread significantly
in width, then the width and the velocity dispersion \cite[\S 8.3.3]{gd2}
are approximately related by,
\begin{equation}
{\sigma_{v\perp} \over v} \simeq {w \over \rperi} = {r\angwid \over \rperi},
\end{equation}
where $w$ is the physical width of the stream, and $\rperi$ is the
radius of the stream's perigalacticon. This gives,
\begin{equation}
\sigma_{\alpha,v} = {r \angwid \over \rperi \sin \beta}.
\label{eq:alpha:equiv}
\end{equation}
If secular
spread has made the stream become wider over time, then 
this relation will over-estimate $\sigma_{\alpha,v}$, since
$\sigma_{v\perp}/v$ is roughly constant.
$\angwid$ therefore represents
an upper bound on the true value of $\sigma_{\alpha,v}$ through this relation.
 This argument also 
assumes that the stream was created from its progenitor in a single
tidal event. Real streams do not form in this way. However, repeated
tidal disruptions can be viewed as a superposition of ever younger streams,
created from a progenitor of ever smaller $\sigma_{v\perp}$.
\eqref{eq:alpha:equiv} holds for each of these individually.
Thus, $\angwid$ remains a good upper bound for $\sigma_{\alpha,v}$
through this relation.

In reality, we do not measure the proper motion of individual
stream stars, but rather the mean motion of a field of $N$ stars. 
The contribution to $\sigma_\alpha$ is from the error on this mean.
Putting this together
with \eqref{eq:errora:curvefitting} gives our final expression
for $\sigma_\alpha$,
\begin{equation}
\sigma^2_\alpha = {\sigma^2_{\alpha,v}\over N} + \sigma^2_{\alpha,f} = { r^2 \angwid^2 \over N
\rperi^2 \sin^2\beta} + {4\angwid \over \anglen^2}.\label{eq:errora:random}
\end{equation}
We note that the first term represents a random error, and the second
term represents a systematic error that will vary with position down
the stream. In general, $\sin \beta$ and $\rperi$ are a priori
unknown. We can infer $\sin \beta$ from radial velocity information,
either directly where the measurements exist, or indirectly from
Galactic parallax distances. Guessing $\rperi$ requires assumptions
to be made about the dynamics, but in general we expect the ratio $r/\rperi \simeq 1$
or less.

Explicit evaluation of $\sin \beta$
and $\rperi$ are not necessary to evaluate the uncertainty if $\sigma_\alpha$
is dominated by the error from fitting, $\sigma_{\alpha,f}$.
We can see this will be the case when the number of observed stars
per field,
\begin{equation}
N > \Big( { r \anglen \over 2 \rperi \sin \beta} \Big)^2.
\label{eq:condition}
\end{equation}
We expect this to be true in almost all practical cases.

\subsection{Uncertainty in tangential velocity calculations}

\eqref{eq:fundamental} can also be used to solve for $\dot{u}$,
\begin{equation}
\dot{u} = {\mu (t_y + t_x) + \Pi (\voty + \votx) \over \cos \alpha
+ \sin \alpha},\label{eq:udot1}
\end{equation}
which becomes,
\begin{equation}
\dot{u} = \mu (\tx + \ty)  + \Pi (\vpar + \vper) = \mu \tx + \Pi \vpar,
\label{eq:udot2}
\end{equation}
when we set $\alpha = 0$. \eqref{eq:udot1} combined with
\eqref{eq:pi1} can be used to explicitly propagate
uncertainties in the measured quantities to $\dot{u}$. 
When $\alpha = 0$, the uncertainty in $\dot{u}$ is,
\begin{eqnarray}
{\sigma^2_{\dot{u}} \over \dot{u}^2} &=&
{\vot^2 \smt^2 \over \mu^2 (\ty \vpar - \tx \vper)^2}
+ { \ty^2 (\vper^2 \svpar^2 + \vpar^2 \svper^2)
\over \vper^2 (\ty \vpar - \tx \vper)^2} \nonumber\\
&&
{}-{ 2 \ty^2 \vpar \vper \cov(\vpar, \vper)
\over \vper^2 (\ty \vpar - \tx \vper)^2}
+ {\vpar^2 \sigma_\alpha^2 \over \vper^2}.
\label{eq:error:udot}
\end{eqnarray}
$\sigma_{\vpar}$ and $\cov(\vpar,\vper)$ are calculated directly from
the error ellipsoid on $\vect{v}_0$, which we have assumed known.

\subsection{Practicality of Galactic parallax as a distance measuring tool}

Using \eqref{eq:pi2} to eliminate $\mu t_\perp$ from
\eqref{eq:error:pi}, and taking the dot product of $\vecthat{p}$ with
\eqref{eq:fundamental} to simplify the last term, we obtain,
\begin{eqnarray}
{\sigma^2_\Pi \over \Pi^2} &=& {1 \over \vper^2} \Big\{
(r \smt)^2 + 
\sigma_{\vper}^2
+ (r\dot{u})^2\sigma_\alpha^2 \Big\}\nonumber\\
& =& {1 \over \vper^2} \Big\{
(r \smt)^2 + 
\sigma_{\vper}^2
+ v^2 \left({r^2 \angwid^2 \over \rperi^2 N} + {4\angwid^2 \over \lendeproj^2}\right)
\Big\},
\label{eq:error3}
\end{eqnarray}
where we have noted that $r\dot{u} = v_t = v\sin\beta$, and we have related
the observed stream length, $\anglen$, to the deprojected length,
$\anglen = \lendeproj \sin \beta$. We note that the last
term is independent of $r$, since $r \angwid = w$ and $\angwid/\lendeproj$
are both constant,  and that for a stream of given physical dimension,
the uncertainty in $\Pi$ has no dependence upon
the angle of the stream $\beta$ to the line of sight.

What level of uncertainty does \eqref{eq:error3} predict, when
realistic measurement errors are introduced? The answer to this is
dependent upon the both the physical properties of the stream $(R_p, \angwid,
\lendeproj, v)$ and the geometry of the problem in question $(r, \vper)$.

We progress by assuming `typical' values for some of these quantities.
The average magnitude of $\vot$ taken over the whole sky is $v_0\, \pi
/ 4$. The average perpendicular component, for a randomly oriented
stream, is $2 / \pi$ of this value.  We therefore assume a typical
value for $\vper$ of $v_0 / 2 \sim 120 \kms$. We also assume a typical
grf velocity equal to the circular velocity, $v = v_c \sim 220\kms$.

\cite{mb09} recently summarised the current state of knowledge of $\vo$. The
uncertainty quoted is typically $\near 5$ percent on each of $(U,V,W)$.
Correspondingly, we estimate a typical value for the uncertainty
$\sigma_{\vper}$ of 5 percent of $\vper$, or $6\kms$.

The GD-1 stream that we
consider below is exceptionally thin and long, with $\angwid \near
0.1\deg$ and $\anglen \near 60\deg$.  The Orphan stream
\citep{gostream,ostream} is of similar length, but about 10 times thicker. Both
of these streams are near apsis, so $\anglen = \lendeproj$. We
therefore take $\angwid \near 1\deg$, $\lendeproj \sim 60\deg$ as
typical of the streams to which one would apply this method.
If hundreds of stars are observed for each proper
motion datum, then \eqref{eq:condition} is true for all realistic
combinations of $(r,\rperi)$, so we can ignore the contribution of
$\sigma_{\alpha,v}$ to $\sigma_\alpha$. The contribution from
$\sigma_{\alpha,f}$ gives $\sigma_{\alpha} \simeq 1.9\deg$.

The individual USNO/SDSS proper motions \citep{munn} used by
\krh have a random uncertainty $\smt\sim 4\masyr$. After averaging over
hundreds of stars and accounting for a contribution from non-stream stars,
\krh report a random uncertainty of $\smt \sim 1 \masyr$ on their GD-1 data.
For a stream $10\kpc$ distant, with these proper motions and the typical values mentioned,
 \eqref{eq:error3} reports an
uncertainty of $\sigma_\Pi / \Pi \sim 40$ percent. By far the
greatest contribution comes from the first term in \eqref{eq:error3},
hence, the error on proper motion measurement is dominating our
uncertainty.

To obtain an uncertainty of $\sigma_\Pi / \Pi < 20$ percent with
\cite{munn} proper motion measurements, we would need to restrict
ourselves to streams less than $5\kpc$ distant. $20$ percent error is
also possible at $10\kpc$ given optimum problem geometry. This is
clearly competitive with the $\sim 20\pc$ at which one could observe a
standard trigonometric parallax, with similar accuracy, using
astrometry of this quality. However, previous work
\citep[\krh]{willett} shows that SDSS photometry combined with
population models produce distance estimates accurate to $\sim 10$
percent for stars in streams at $8\kpc$.  The accuracy of Galactic
parallax is therefore not likely to be as good as that of photometric
distances for distant streams, using data this poor, unless the
problem geometry is favourable.

Proper-motion data from the Pan-STARRS telescope
is expected to be accurate to $\near 1\masyr$ for Sun-like stars at
$10\kpc$ \citep{pan-starrs-3pi}. \krh reduce raw data with accuracy
$\near 4\masyr$ to processed data accurate to $\near 1\masyr$, even
though the expected proper motion of the stars is of the same size as
the errors. It is not unreasonable to expect a similar analysis
applied to Pan-STARRS raw data, where the relative error would be much
less than unity, to yield processed data accurate to $\sim
0.2\masyr$. In truth, the ability of Pan-STARRS to detect very
faint stars will increase the number of stars identifiable with a stream,
and thus reduce the uncertainty in the mean proper motion further than this,
but we use $0.2\masyr$ as a conservative estimate.

The same $10\kpc$ distant stream would have a parallax error of
$\sigma_\Pi / \Pi \simeq 11$ percent with data this accurate.  An
error of less than $20$ percent is possible for a typical stream less
than $\sim 23\kpc$ distant, and for a stream with favourable geometry
less than $50\kpc$ distant. \cite{juric} report that SDSS photometric
distances for dwarf stars have $\near 40$ percent error at
$20\kpc$. Thus, the accuracy of Galactic parallax derived from
Pan-STARRS data should be at least comparable to distance estimates
from photometric methods, even in the typical case.

Future projects such as LSST and Gaia will each obtain proper motions
accurate to $\sim 0.2\masyr$ for Sun-like stars $10\kpc$ distant \citep{ivezic,
gaia}. These
data would allow a distance estimate for our typical stream accurate
to $8$ percent, and a stream with favourable geometry accurate to
$4$ percent. Error in the proper motion no longer dominates the
uncertainty in these calculations.  We might expect such accurate
astrometric surveys to reduce the uncertainty in the Solar motion; in
this case, the error in parallax would be lower still.

Gaia will not observe Sun-like stars beyond $10\kpc$, but LSST will, with
accuracy of $0.4\masyr$ for dwarf stars $30\kpc$ distant
\citep{ivezic}. The accuracy of the parallax to our typical stream at
this distance would be about $14$ percent with these data, and
$6$ percent is achievable with optimum geometry. A typical stream
could be measured to 20 percent accuracy out to $40\kpc$, and a stream
with favourable geometry out to $54\kpc$; this range approaches the
limit of LSST's capability for detection of dwarf stars.  Such data
will put the Orphan stream, which is about $20-30\kpc$ distant
\citep{gostream, ostream, sales-orphan}, in range of accurate trigonometric
distance estimation. For comparison, photometric distances from SDSS
data are hardly more accurate than 50 percent for this stream
\citep{ostream}.

\begin{figure}
\centerline{\epsfig{file=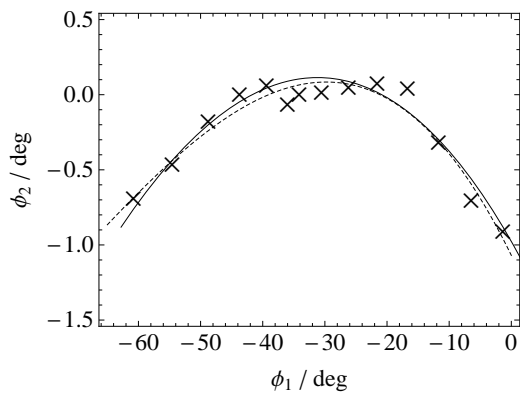,width=.8\hsize}}
\caption{Full line: the orbit for the GD-1 stream,
taken from \krh. Crosses: pseudo-data
derived from that orbit, but randomly scattered in $\phi_2$
according to a Gaussian distribution with a dispersion of $\sigma_{\phi_2} = 0.1\deg$.
Dotted line: a cubic polynomial fitted to the pseudo-data,
used to estimate stream direction.}
\label{fig:pseudodata}
\end{figure}

\begin{figure}
\centerline{\epsfig{file=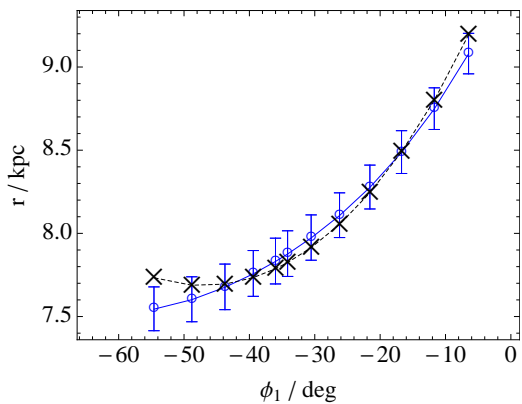,width=.8\hsize}}
\caption{Dotted line: the orbit of the GD-1 stream,
taken from \krh. Crosses: the true distance of each
pseudo-datum. Circles: Galactic parallax
distances computed from the pseudo-data.
The error bars represent
 the distance error expected from the polynomial fitting
 procedure.
No extraneous error was added. The estimated errors
are show to be a good estimate of likely error from
the fitting procedure, and the agreement of the
 distances overall is excellent.}
\label{fig:pd-distance}
\end{figure}

\begin{figure}
\centerline{\epsfig{file=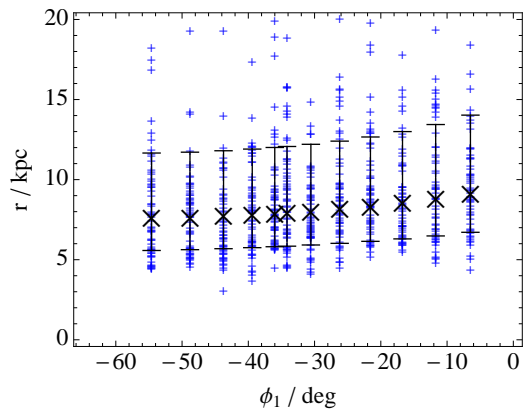,width=.8\hsize}}
\caption{Crosses: Galactic parallax distances
computed from the pseudo-data, with no extraneous errors.
Error bars: the random scatter expected in Galactic
parallax distances, with measurement errors
 as mentioned in the text. Plus signs: Galactic parallax
distances computed from 60 Monte Carlo realisations of each
pseudo-datum convolved with the measurement errors. The analytic
uncertainty estimate and the Monte Carlo realisations are
in good agreement.}
\label{fig:pd-MC}
\end{figure}

\section{Tests}\label{sec:tests}

To test the method, pseudo-data was prepared from an orbit fitted to
data for the GD-1 stream by \krh. The orbit is described by
the initial conditions $\vect{x}=(-3.41,13.00,9.58) \kpc$, $\vect{v} =
(-200.4,-162.6,13.9) \kms$, where the $x$-axis points towards the
Galactic centre, and the $y$-axis points in the direction of Galactic
rotation. The orbit was integrated in the logarithmic
potential,
\begin{equation}
\Phi(x,y,z) = {v_c^2 \over 2} \log \left(
x^2 + y^2 + \left( { z \over q } \right)^2 \right),
\end{equation}
where $v_c = 220\kms$ and $q = 0.9$. The resulting trajectory was
projected onto the sky, assuming a Solar radius
$R_0 = 8.5 \kpc$.  Several points were sampled, and each was taken to be a
separate datum in the pseudo-data set. The proper motion
for each datum was computed by projecting the difference 
between its grf motion and the Solar motion on to the sky.

The pseudo-data were transformed into the rotated coordinate system
used by \krh to facilitate comparison with their data; the
transformation rule is given in the appendix to \krh. The stream is very
flat in this coordinate system, so the dependence of $\phi_2$ on
$\phi_1$ is relatively weak. This helps to increase the quality of the fitted curve
and minimises the corresponding error in $\sigma_{\alpha,f}$.

To simulate the observed scatter in the real positional data, the
pseudo-data were each scattered in the $\phi_2$ coordinate according
to a randomly-sampled Gaussian distribution with a dispersion
$\sigma_{\phi_2} = 0.1\deg$. The resulting positional pseudo-data are
plotted in \figref{fig:pseudodata}, along with the orbit from which
they were derived (full curve). A cubic polynomial representing $\phi_2(\phi_1)$ was
least-squares fitted to the pseudo-data, the tangent of which was used to estimate
$\vecthat{p}$.  In the case of the pseudo-data, uniform
weights were applied to each datum for the fitting processes. The
resulting curve is also show in \figref{fig:pseudodata} (dotted curve).

When the correct orbit is
used to calculate $\vecthat{p}$, and precise values for the measured
proper motion $\mu \vecthat{t}$ and Solar reflex motion $\vot$ are
used, the distance is recovered perfectly from \eqref{eq:pi2}.
\figref{fig:pd-distance} compares the recovered distance when
$\vecthat{p}$ is estimated using the polynomial fit to the
pseudo-data, but still using accurate values for $\mu \vecthat{t}$ and
$\vot$. Our pseudo-data stream is $\angwid \simeq 0.1\deg$ wide and
$\anglen \simeq 60\deg$ long. \eqref{eq:errora:curvefitting}
therefore estimates $\sigma_{\alpha,f} \simeq 0.38\deg$. The recovered
distances in \figref{fig:pd-distance} are in error by only $\sim 2$
percent across most of the range, which is the approximate uncertainty
predicted by \eqref{eq:error:pi} for this value of
$\sigma_{\alpha,f}$. Thus, the estimation of $\vecthat{p}$
from the observed stream is good, and contributes little error to the
distance calculations.

The \krh observational data for the GD-1 stream, discussed below, have a similar
uncertainty $\sigma_\alpha \sim 0.38\deg$ due entirely to the fitting process,
and proper motion uncertainties $\sigma_\mu \sim 1\masyr$.
\figref{fig:pd-MC} shows the recovered distances from
\figref{fig:pd-distance} with error bars for the expected uncertainty
in recovered distance, given these
measurement uncertainties and the uncertainty in $\vo$ quoted
in Section 1.  Also plotted for each datum are the
distances recovered from 60 Monte Carlo realisations of the
pseudo-data input values, convolved with the errors given above.

\eqref{eq:error:pi} is found to be a good estimator
for the uncertainty, with approximately 80 percent of
the Monte Carlo realisations falling within the error bars.
The error in parallax for the \krh data is thus predicted
to be about 50 percent, of which the greatest contribution
comes from the uncertainty in proper motion.

\begin{figure}
\centerline{\epsfig{file=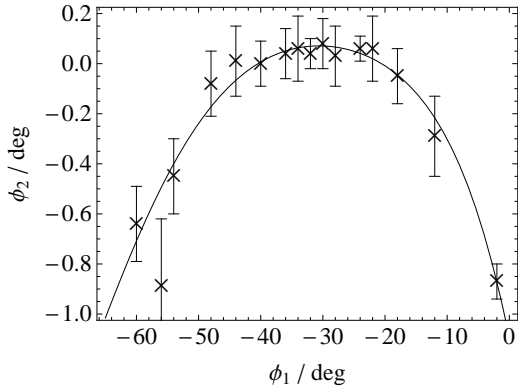,width=.8\hsize}}
\caption{Crosses: on-sky position data for the GD-1
stream, as published in \krh. The error bars
represent the quoted uncertainties. Full line:
linear least-squares fit of a cubic polynomial, $\phi_2(\phi_1)$,
to these data; the inverse-square of the uncertainties
was used to weight the fit.}
\label{fig:gd1-fit}
\end{figure}

\begin{figure}
\centerline{\epsfig{file=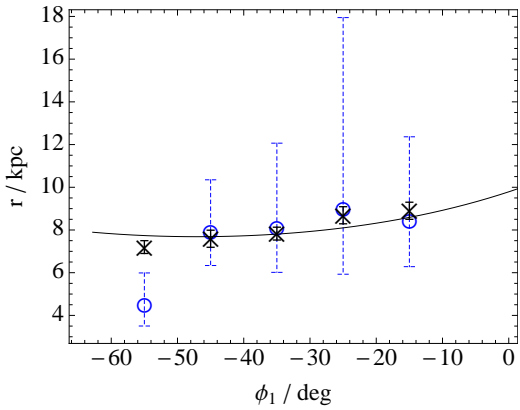,width=.8\hsize}}
\caption{Circles: Galactic parallax distances for the GD-1 data
  presented in \krh. Dotted error bars: the uncertainty estimated by
  \eqref{eq:error:pi}, given the \krh measurement
  uncertainties. Crosses: the photometric distances reported in \krh,
  along with their error bars. Full line: the orbit for GD-1 taken
  from \krh.  With the exception of the datum at $\phi_1 \sim -55$
  deg, the Galactic parallax distances are in excellent agreement with
  the photometric distances from \krh. The dotted error bars appear to
  seriously over-estimate the true error in the distance estimates.
}\label{fig:gd1-dist}
\end{figure}

\begin{figure}
\centerline{\epsfig{file=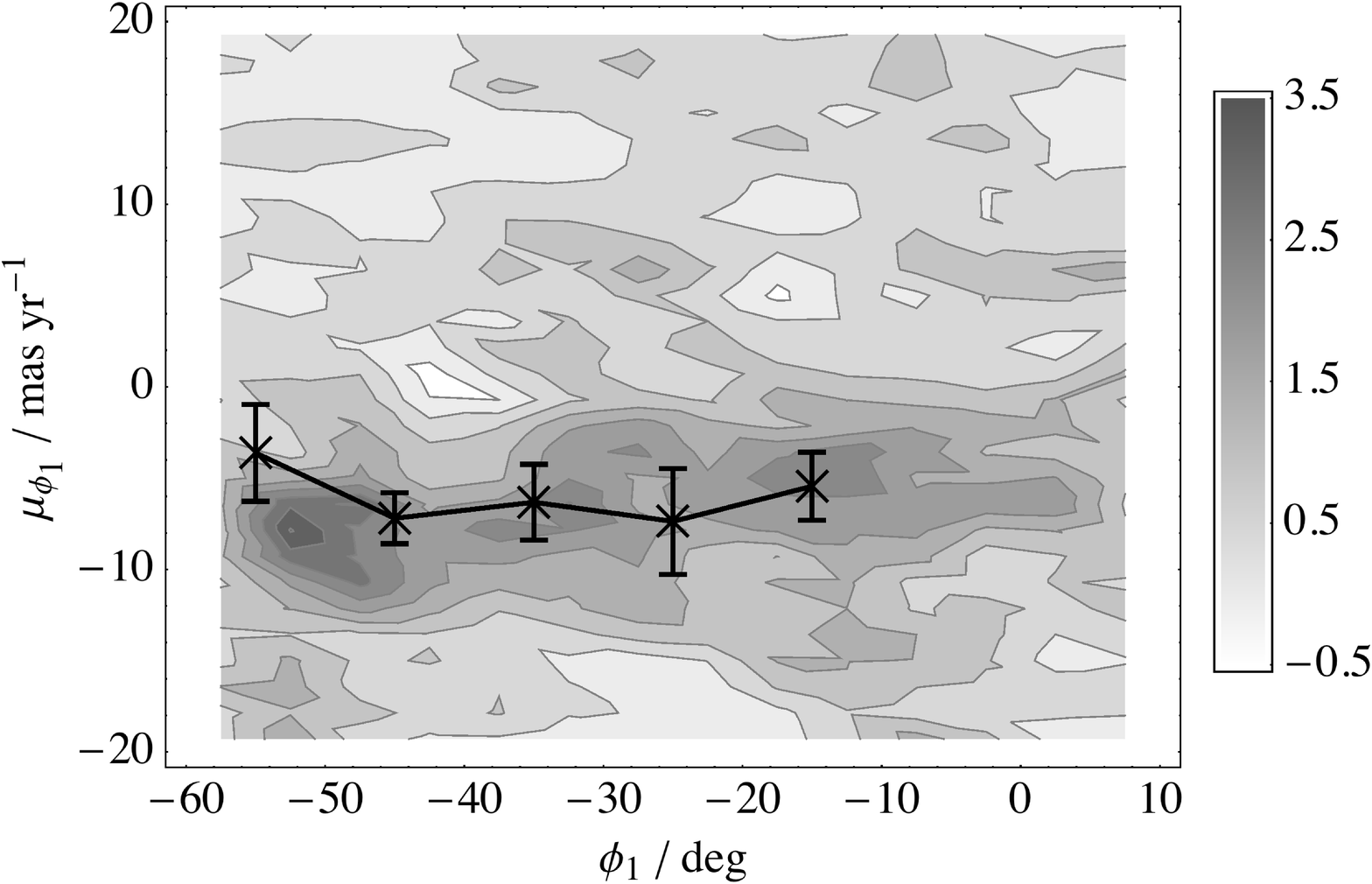,width=.95\hsize}}
\centerline{\epsfig{file=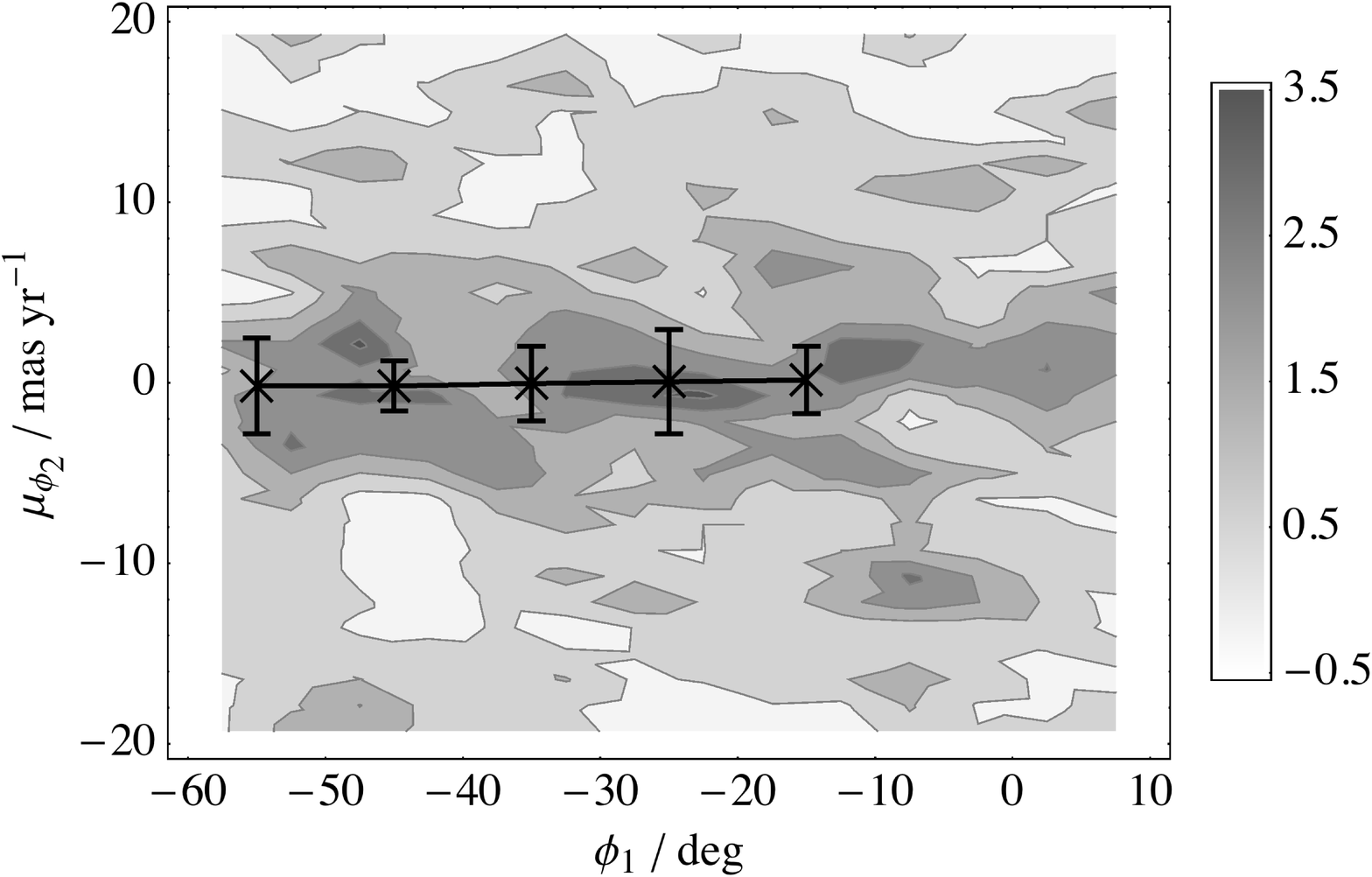,width=.95\hsize}}
\caption{Full lines: Galactic-rest frame proper motion
($\dot{u}$) calculated from the \krh data using
\eqref{eq:udot2}. The (upper, lower) panels show
the $(\phi_1, \phi_2)$ components respectively. Plotted in the background
are the observational data from Fig. 9 in \krh;
the greyscale shows the number of stream stars, per bin, with the
given motion. 
The data are broadly consistent,
except for the datum at $\phi_1 \sim -55$ deg in the upper panel.
}\label{fig:gd1-udot}
\end{figure}

\section{Distance to the GD-1 stream}\label{sec:gd1}

\figref{fig:gd1-fit} shows the on-sky position
data for the GD-1 stream, as published in \krh.
Also shown in \figref{fig:gd1-fit} is a
linear least-squares fit of a cubic polynomial
to these data, used to estimate $\vecthat{p}$. The weights for the fit
were the inverse-square uncertainties for each position field, as given
by \krh.

\krh provide measured proper motion data for five fields
of stars, spanning the range $\phi_1 \sim (-55, -15)\deg$,
along with uncertainties for these measurements.
Uncertainty in the stream direction is $\sigma_\alpha \sim 0.38\deg$,
which is entirely contributed by the curve
fit to the stream; since hundreds of stars contributed to the
calculation of the proper motions, the contribution from
the first term in \eqref{eq:errora:random} is negligible. The uncertainty in
$\vot$ is computed for each individual field from the
uncertainty in $\vo$ given in Section 1.

\figref{fig:gd1-dist} shows the Galactic parallax distances for each of these
data, along with the \krh photometric distances. The dotted error bars
represent the expected error in distance for
the uncertainties given. The small solid error bars are the uncertainties
reported by \krh for their photometric distances. The \krh orbit used to
compute the earlier pseudo-data is plotted for comparison.

With the exception of the datum at $\phi_1 \sim -55 \deg$, the
parallax distances and the \krh distances are in remarkable
agreement. However, the dotted error bars vastly overestimate the true
error in the results.  If we ignore the datum at $\phi_1 \sim -55
\deg$, the scatter in the distance, $\sigma_r \sim 1\kpc$, is similar
to that of the photometric distances, and consistent with a true
random error of $\sigma_\mu \sim 0.3\masyr$, and negligible systematic
offset. We cannot explain this discrepancy, except by suggesting that
the \krh proper motion measurements are more accurate than the
published uncertainties suggest. This is corroborated by the top-right
panel of Fig. 13 from \krh in which the $\mu_{\phi_2}$ data, with the
exception of the datum at $\phi_1 \sim -55 \deg$, show remarkably
little scatter within their error bars.

\figref{fig:gd1-udot} shows the Galactic rest-frame proper motions,
$\dot{u}$, calculated from \eqref{eq:udot2} along with their error bars,
from \eqref{eq:error:udot}.  In the background are plotted the data
from Fig. 9 of \krh, which show the density of stars with a given grf
proper motion in the sample of stars chosen to be candidate members of
the stream, and after subtraction of a background field.  The \krh grf
proper motions have been calculated by correcting measured proper
motion for the solar reflex motion, using an assumed distance of
$8\kpc$ (Koposov, private communication); this assumption will cause a
systematic error in the \krh proper motions, of order the distance
error, which changes with position down the stream. The apparently
large width of the stream in this plot is due to uncertainty in the
underlying \cite{munn} proper motion data.

The stream is clearly visible in this plot as the region of high
density spanning $\phi_1 \sim (0, -60) \deg$ with $\mu_{\phi_2}
\simeq 0 \masyr$ and $\mu_{\phi_1}$ falling slowly between
$(-6, -10 )\masyr$. Despite the
expected systematic error, the estimates of
$\dot{u}$ from the parallax calculation are consistent with these
data, with the exception of the same datum at $\phi_1 \sim -55 \deg$
that also reports an anomalous distance.

We explain this suspect datum as follows. From inspection of the
top-right panel of Fig. 13 from \krh, it is apparent that the
$\mu_{\phi_2}$ measurement for this datum is not in keeping with the
trend. Conversely, the corresponding $\mu_{\phi_1}$ measurement is not
obviously in error. If the magnitude of $\mu_{\phi_2}$ for this datum
has been over-estimated by the \krh analysis, then \eqref{eq:pi2} will
over-estimate the parallax, and hence under-report the distance.
\figref{fig:gd1-dist} indicates that the distance for this datum
is indeed under-reported.

The effect of such an error in $\mu_{\phi_2}$ on the
grf proper-motion, $\dot{u}$, can be understood by
considering \eqref{eq:udot2}. If $\Pi$ is over-estimated,
$\dot{u}$ will be either over-estimated or under-estimated, depending
on the relative sign of the two terms. In the case of GD-1,
$\mu \tx$ and $\vpar$ have opposite signs, so an over-estimated
$\Pi$ will result in an under-estimated $\dot{u}$. This 
too corresponds with the behaviour of the suspect datum in
\figref{fig:gd1-udot}.

It is unknown why this particular datum should be significantly in
error while the other data are not. There are no obvious structures in
the lower panel of \figref{fig:gd1-udot} which might cause the fitting
algorithm in \krh to mistakenly return an incorrect value for
$\mu_{\phi_2}$. Nonetheless, if the scatter in the other data are
accepted as indicative of their true statistical error, it is clear
that the datum at $\phi_1 \sim -55\deg$ cannot represent the proper
motions of GD-1 stars at that location.
We therefore predict that an appropriate
re-analysis of the proper-motion data, taking care to ensure that 
a signal from GD-1 stream stars is properly detected, will return a revised
proper-motion of $\mu_{\phi_2} \sim -3 \masyr$.

In summary, it seems that Galactic parallax measurements confirm the
\krh photometric analysis, and predict that the stream is
approximately $(8 \pm 1)\kpc$ distant, where the uncertainty denotes
the scatter in the results. Since Galactic parallax and photometric
estimates are fundamentally independent, it seems unlikely that
systematic errors in either would conspire to produce the same shift
in distance; this implies that no systematic error is present.

We also calculate a grf proper motion for the stream of
$\mu_{\phi_1} = (-7 \pm 2) \masyr$, corresponding to a grf tangential
velocity of $(265 \pm 80)\kms$ in a direction $(\mu_l \cos b, \mu_b)
\simeq (0.8, -0.6)$. This implies that the stream is on a retrograde orbit, inclined to
the Galactic plane by $\near 37\deg$, which is in accordance with previous
results \citep{willett,koposov}.

The galactocentric radius of $\near 14.5\kpc$ does not seem to be
changing rapidly along stream's length, which subtends $\near 12\deg$
when viewed from the Galactic centre. This implies that the observed
stream is at an apsis. The grf velocity of the stream is faster than the
circular velocity, $v_c \sim 220\kms$.  This implies that the stream
is at pericentre, although the large uncertainty prevents a firm
conclusion from being drawn. We note that the radial velocity
data in \krh would also imply that the stream is observed at
pericentre.

\section{Conclusions}\label{sec:conclusions}

We have demonstrated the practical application of a technique for
computing Galactic parallax, as described by \paper1.  This
technique utilises the predictable trajectories of stars in a stream
to identify the contribution of the reflex motion of the Sun to the
observed proper motion.  The parallax and the Galactic rest-frame
proper motion follow from this.

The only assumption made is knowledge of the Galactic rest-frame
velocity of the Sun. It is also a requirement that the observed stars
are part of a stream. Recent evidence
\citep{Odenkirchen02,sagdf,yanny,Fieldstars,gostream,GD-1,ngc5466,
  g09,newberg} indicates that tidal streams are a common constituent
of the Galactic halo, and so this technique should have widespread
application.

We have derived an expression for the uncertainty in
Galactic parallax calculations. We include contributions
from measurement errors in proper motion and Solar motion,
error in the estimation of stellar trajectories from the
stream direction, and algorithmic error in the estimation of
stream direction itself.

The uncertainty for calculations involving a particular stream is
depend upon the size, location and orientation of the stream, as well
as upon measurement errors. We estimate that using individual proper
motions accurate to $4\masyr$, available now in published surveys
\citep{munn}, the parallax of a $10\kpc$ distant stream with typical
geometry could be measured with an uncertainty of 40
percent. The parallax of a stream with optimum geometry could be 
measured with approximately half this uncertainty. 

Proper motion data
from the forthcoming Pan-STARRS PS-1 survey \citep{panstarrs,pan-starrs-3pi} will yield the
distance to a typical $10\kpc$ distant stream with 11 percent accuracy, or
the distance to a stream at $23\kpc$ with 20 percent accuracy; with favourable
geometry this accuracy could be achieved for a stream as distant
as $50\kpc$. With data of this
quality, the uncertainty in distances from Galactic parallaxes
will be considerably lower than those of photometry for distant
streams.

The LSST \citep{lsst,ivezic} and the Gaia
mission \citep{gaia} will produce proper-motion data that are more
accurate still.  Such data would allow the distance to stars in a
$10\kpc$ typical stream to be computed to an accuracy of order of 8
percent, where the limitation is now imposed by uncertainty in the
solar motion and in the stream trajectory. It is likely that LSST and
Gaia data will allow the uncertainty in the solar motion to be
significantly reduced, so in reality much better precision can be
expected at this distance.  For streams $30\kpc$ distant, LSST proper
motions will allow distance estimates as accurate as 14 percent to be
made in the typical case, and 6 percent with optimum geometry. Thus,
the high-quality astrometric data that is expected to be available in
the next decade will allow parallax estimates for very distant streams
to be made with unparalleled accuracy.

To test the method presented, we have created
pseudodata simulating the GD-1
stream \citep{GD-1}. When the
method is provided with error-free pseudo-data,
the correct parallax is computed perfectly.
When errors are introduced into the pseudo-data,
the reported parallax degrades in line with
the uncertainty estimates.

We applied the method to the astrometric data for the GD-1
stream in \cite{koposov}. With the exception of a single datum, the
Galactic parallax is remarkably consistent with the photometric
distances quoted by \cite{koposov}.
Indeed, the uncertainty in the measured proper motions quoted by
\cite{koposov} should produce significant error in the Galactic
parallax. However, the scatter in the results is consistent
a random error of only $\sim 0.3\masyr$, and if the photometric
distances of \krh are believed, no systematic offset.
This is at odds with the typical uncertainty in
the proper motion of $\sim 1\masyr$ reported by \cite{koposov}.  We
cannot explain this discrepancy, other than to suggest that the
\cite{koposov} method for estimating error in the proper motions is
producing significant over-estimates.

The Galactic rest-frame proper motions predicted for the stream are
also consistent with observational data from \cite{koposov}, with the
exception of the same datum that also reports an inconsistent distance.  We
conclude that the proper-motion associated with this datum is erroneous,
and we predict that reanalysis of the stream stars near this datum
will reveal a reduced proper-motion measurement of $\mu_{\phi_2} \sim 3\masyr$.

Photometry and Galactic parallax produce fundamentally independent
estimates of distance. The quality of the corroboration of the
\cite{koposov} photometric distance estimates for GD-1 by the Galactic
parallax estimates presented here therefore lends weight to the
conclusion that the predicted distance, in both cases, is correct.  On
this basis, we conclude that the GD-1 stream is about $(8 \pm 1) \kpc$
distant from the Sun, on a retrograde orbit that is inclined $37\deg$
to the Galactic plane with a rest-frame velocity of $(265 \pm
75)\kms$. We also conclude that the visible portion of the stream
is probably at pericentre.

The prospect of being able to map trigonometric distances in the Galaxy to high
accuracy at tens of kiloparsecs range is indeed exciting.  The 
distances generated using this method, although limited to stars in
streams, could be used to calibrate other distance measuring tools,
such as photometry, that would be more widely applicable. The technique is
immediately applicable to any stream for which proper-motion data are
currently available, although we anticipate limited accuracy until better
proper-motion data are available.

Given enough parallax data points along a given stream, an orbit can
be constructed by connecting those points. This orbit is predicted
independently of any assumption about the Galactic potential, which it
must strongly constrain. Constraints on the Galactic potential impose
constraints on theories of galaxy formation and cosmology. It would
seem that the combination of dynamics and Galaxy-scale precision
astrometry, such as provided by this method, could well have profound
implications for astrophysics in the future.  At present, however, it
is not obvious how to combine all sources of astrometric and dynamical
information, to produce the tightest constraints on the potential. We therefore
encourage the exploration of methods for combining this information,
in anticipation of the arrival of higher quality astrometric data in the next
few years.

\section*{Acknowledgments}

I thank Prof. James Binney for his insight and
continued support during this work; and the Oxford Dynamics group for
their helpful remarks. I also thank Prof. Andy Gould for his thoughts
on uncertainties which provoked the analysis in section 2, 
Sergey Koposov for his provision of the data for \figref{fig:gd1-udot},
and the anonymous referee for his/her remarks. I
acknowledge the support of PPARC/STFC during preparation of this work.

\label{lastpage}

\end{document}